\documentclass[a4paper,11pt]{article}

\usepackage{jheppub}
\usepackage{graphicx}
\usepackage{amsmath}
\usepackage{slashed}
\usepackage{braket}
\usepackage[export]{adjustbox}
\usepackage{enumitem}
\usepackage{placeins}
\usepackage[normalem]{ulem}
\usepackage{multirow}

\newcommand{\eq}[1]{eq.~\eqref{eq:#1}}

\newcommand{\fig}[1]{figure~\ref{fig:#1}}

\newcommand{\df}{\mathrm{d}}

\newcommand{\eps}{\epsilon}

\newcommand{\bn}{{\bar n}}

\newcommand{\GammaC}{\Gamma_{\rm cusp}}
\newcommand{\nn}{\nonumber}
\newcommand{\MSbar}{\overline{\mathrm{MS}}}

\newcommand{\fth}{f^{\rm{th}}}

\newcommand{\Sth}{S_{\rm{thr.}}^r}

\newcommand{\tr}{{\rm tr}}

\def\beq{\begin{equation}}
\def\eeq{\end{equation}}
\def\bea{\begin{eqnarray}}
\def\eea{\end{eqnarray}}

\title{\boldmath Soft Integrals and Soft Anomalous Dimensions at N$^3$LO and Beyond}

\author[a]{Claude Duhr,}
\emailAdd{cduhr@uni-bonn.de}
\author[b]{Bernhard Mistlberger,}
\emailAdd{bernhard.mistlberger@gmail.com}
\author[b]{Gherardo Vita,}
\emailAdd{gherardo@slac.stanford.edu}

\affiliation[a]{Bethe Center for Theoretical Physics, Universit\"at Bonn, D-53115, Germany}
\affiliation[b]{SLAC National Accelerator Laboratory, Stanford University, Stanford, CA 94039, USA}

\abstract{
We calculate soft phase-space and loop master integrals tor the computation of color-singlet cross sections through N$^3$LO in perturbative QCD. 
Our results are functions of homogeneous transcendental weight and include the first nine terms in the expansion in the dimensional regulator $\epsilon$. 
We discuss the application of our results to the computation of deeply-inelastic scattering and $e^+e^-$ annihilation processes.
We use these results to compute the perturbative coefficient functions for the Drell-Yan and gluon-fusion Higgs boson production cross sections to higher orders in $\epsilon$ through N$^3$LO in QCD in the limit where only soft partons are produced on top of the colorless final state.
Furthermore, we extract the anomalous dimension of the inclusive threshold soft function and of the $N$-Jettiness beam and jet functions to N$^4$LO in perturbative QCD.
}

\preprint{BONN-TH-2022-09, SLAC-PUB-17677}

\begin{document}

\maketitle


\section{Introduction}
\label{sec:intro}
Analytic computations for scattering cross sections play a crucial role in the field of high energy particle physics phenomenology. 
In such computations we use perturbative quantum field theory (QFT) in order to achieve precise predictions for observables that allow us to study the interactions of fundamental particles.
The field of analytic computations is advancing rapidly and has produced many cutting edge results, like predictions for production cross sections to third order in QCD perturbation theory for the Large Hadron Collider(LHC)~\cite{Anastasiou:2015ema,Duhr:2019kwi,Duhr:2020sdp,Duhr:2020seh,Dulat:2018bfe,Mistlberger:2018etf,Duhr:2021vwj}, fourth order QCD results for the production of hadrons in electron-positron collisions~\cite{Herzog:2017dtz,Baikov:2012er} or analytic formulae for event shape observables like the energy-energy correlation function~\cite{Ebert:2020sfi,Dixon:2018qgp,Dixon:2019uzg}.
On top of predictions for explicit cross sections, analytic results play a crucial role to determine many universal quantities appearing as ingredients to the calculation of scattering cross sections.
Shining examples are the splitting functions at third order in perturbative QCD~\cite{Ablinger:2014nga,Ablinger:2017tan,Blumlein:2021enk,Moch:2004pa,Vogt:2004mw}, the so-called cusp anomalous dimension at fourth loop order~\cite{Henn:2019swt,vonManteuffel:2020vjv}, or extraction of the universal infrared behavior of scattering amplitudes at three loop order~\cite{Almelid:2017qju,Almelid:2015jia}.

In this article we discuss and extend a set of analytic ingredients for perturbative computations that have already found widespread application. 
The quantities in question are so-called \emph{soft integrals}. 
These integrals are Feynman integrals for phase-space and loop integrals expanded around a certain kinematic limit - the so-called \emph{soft} or \emph{threshold} limit.
Throughout this article we work within the framework of dimensional regularization. 
Our soft integrals first made their appearance in the computation of hadronic production cross section for a Higgs boson at N$^3$LO~\cite{Anastasiou:2012kq,Anastasiou:2013srw,Anastasiou:2015yha,Anastasiou:2014lda},
where they played a two-fold role: 
First, using the framework of reverse unitarity~\cite{Anastasiou2002,Anastasiou2003,Anastasiou:2002qz,Anastasiou:2003yy,Anastasiou2004a}, it is possible to express the threshold approximation of the production cross section~\cite{Anastasiou:2014vaa,Anastasiou:2014lda} as a linear combination of these soft master integrals.
Second, the soft master integrals served as boundary conditions~\cite{Anastasiou:2013mca,Mistlberger:2018etf} for differential equations~\cite{Kotikov:1990kg,Kotikov:1991hm,Kotikov:1991pm,Gehrmann:1999as,Henn:2013pwa} used to calculate the exact Higgs boson cross section at N$^3$LO in QCD perturbation theory. 

The same soft master integrals were subsequently used in the computation of several analytic results.
First, the computation of the inclusive gluon fusion Higgs boson production cross section at the LHC was extended to the charged current and neutral current Drell-Yan cross sections~\cite{Duhr:2020sdp,Duhr:2020seh,Duhr:2021vwj}, as well as to the production cross section of a Higgs boson from bottom quark fusion~\cite{Duhr:2019kwi,Duhr:2020kzd}. 
In ref.~\cite{Dulat:2017aa} it was realized that analytic computations for more differential quantities, like the rapidity or transverse momentum distributions, can be carried out efficiently thanks to the knowledge of the very same analytic information. 
As a result, it was possible to perform a threshold expansion of differential cross sections for the production of a Higgs boson at N$^3$LO~\cite{Dulat:2017prg} and to compute the rapidity distribution of the Higgs boson~\cite{Dulat:2018bfe} using analytic results.
In ref.~\cite{Ebert:2020lxs} it was pointed out that soft integrals may serve as key analytic ingredients to determine so-called collinear master integrals. 
In turn these results where then used to determine the so-called transverse momentum dependent beam functions at N$^3$LO in QCD~\cite{Ebert:2020yqt} as well as the $N$-jettiness beam functions at the same order~\cite{Ebert:2020unb}.
In ref.~\cite{Ebert:2020qef} it was realized that it is easy to analytically continue the soft integrals computed for a production cross section to serve as ingredients for the computation of a Deep Inelastic Scattering (DIS) process or a cross section relevant for an electron-positron annihilation experiment. As a consequence, the energy-energy correlation function was calculated in the large angle limit at N$^3$LO in QCD perturbation theory~\cite{Ebert:2020sfi}. 
The above results have widespread implication on particle physics phenomenology, which demonstrates the importance of analytic results for soft integrals. 

The soft integrals discussed above were presented in the literature as a Laurent series in the dimensional regulator up to the power required to perform computations at N$^3$LO in QCD perturbation theory. 
Here, we extend this computation to include two additional powers in this Laurent expansion, and consequently we obtain information that will be an ingredient for the computation of scattering cross sections beyond N$^3$LO. 
In particular, we compute two classes of soft integrals: the first one is differential in the four momentum of the color neutral particle, while the second one is integrated over the full final state phase space. 
Our basis of soft integrals is given in terms of pure functions of uniform transcendental weight. 
Here we focus on integrals with two and three partons in the final state as results for single parton final state integrals can be found elsewhere~\cite{Anastasiou:2013mca,Dulat:2014mda,Duhr:2013msa,Duhr:2014nda}.
We then use these new results in order to determine the neutral current Drell-Yan and gluon fusion Higgs boson production cross section in the threshold limit to two orders beyond the finite term in the dimensional regulator at N$^3$LO. 

Threshold factorization~\cite{Sterman:1986aj,Catani:2003zt,Catani:1989ne,Catani:1990rp,Ahrens:2009cxz,Ahrens:2008qu,Ahrens:2010rs} allows one to compute the threshold limit of any colorless production cross section once purely virtual corrections for this process and the so-called threshold soft function are known. 
We compute the threshold soft function through N$^4$LO in perturbative QCD up to one undetermined constant.
We extract explicitly the anomalous dimension of the soft function through N$^4$LO as one of our results.
We find agreement for the threshold anomalous dimension and soft function with the existing results~\cite{Das:2020adl}. 
Building on refs.~\cite{Vogt:2018miu,Das:2020adl,Moch:2017uml,Moch:2021qrk,Moch:2018wjh,Das:2019btv}, as a side product
we are able to determine previously unknown coefficients of the Altarelli-Parisi splitting functions at third non-trivial order.

This article is organized as follows. 
In section~\ref{sec:setup} we introduce our notation and give our definitions of soft phase space and loop integrals.
Next, we discuss our computation of soft loop and phase space integrals for integrals in section~\ref{sec:SoftIntegrals}.
We then apply these soft integrals to the computation of the Drell-Yan and Higgs boson production cross section in the threshold limit through N$^3$LO in perturbative QCD in section~\ref{sec:thresholdxs}.
We generalize our results to generic production cross sections using threshold factorization and extract the threshold soft function and anomalous dimension in section~\ref{sec:fac}.
Finally, we draw our conclusions and summarize our results in section~\ref{sec:conclusions}.
\newpage
\section{Setup}
\label{sec:setup}
 \begin{figure}[!h]
  \begin{center}
  \includegraphics[scale=0.1]{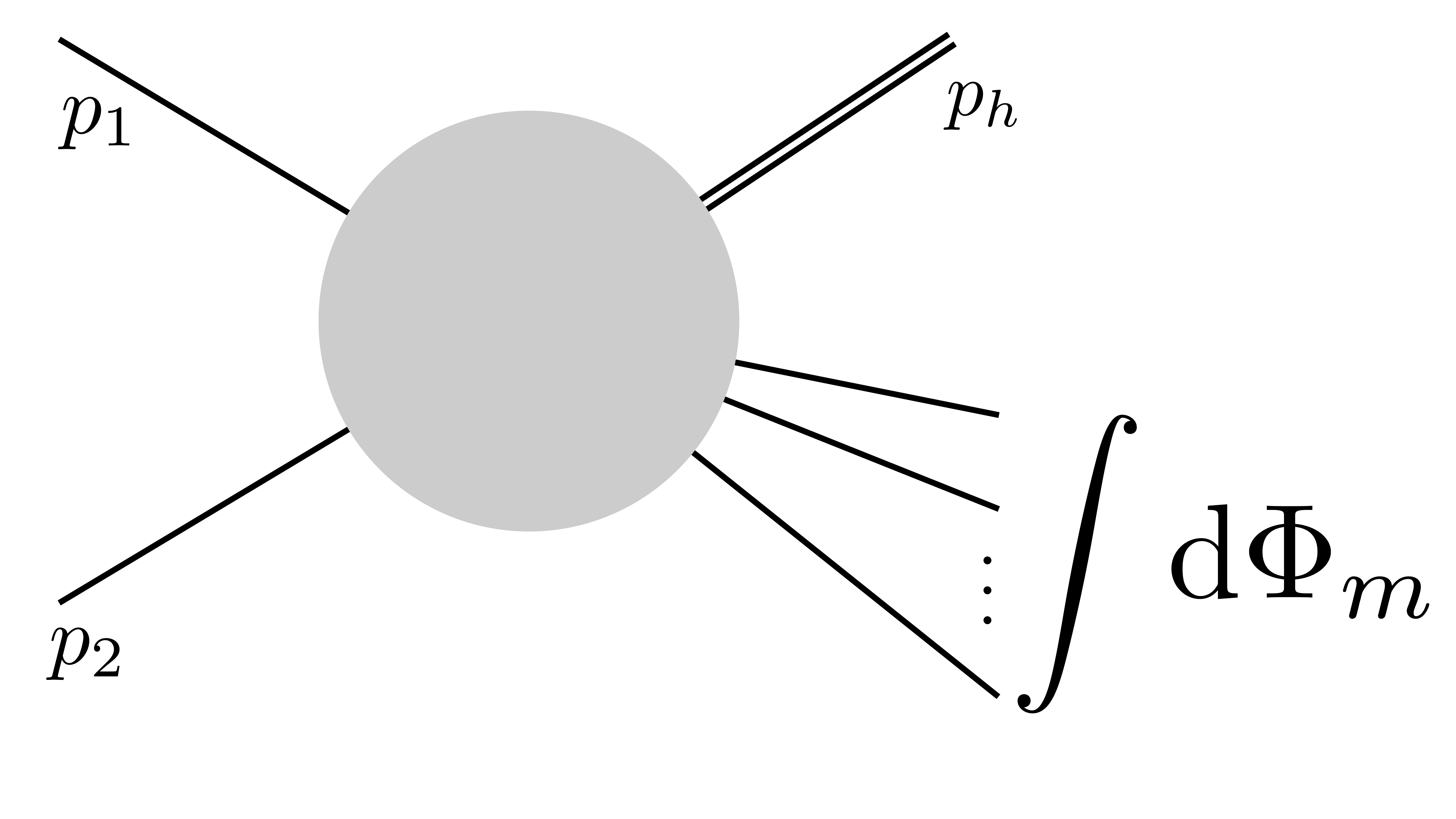}
  \caption{\label{fig:masterform}
Schematic depiction of a partonic scattering cross involving $m$+2 partons and a color neutral particle $h$.
  }
  \end{center}
  \end{figure} 
In this article we discuss Feynman integrals appearing in the computation of scattering cross sections in perturbative QFT. 
We are interested in scattering processes where two partons and one color neutral particle (like a Higgs boson or an off-shell photon) scatter with $m$ massless final state particles.
Schematically, an amplitude for such a process is depicted in fig.~\ref{fig:masterform}. 
We are interested in the case where all kinematic information of these $m$ final state particles is integrated out. 
Defining all momenta to be in-going, momentum conservation is given by
\beq
p_1+p_2+p_h+k=0,
\eeq
where we denote the collective momentum of the $m$ massless final state partons by $k$.
We define the following variables.
\beq
\label{eq:supervars}
s=(p_1+p_2)^2,\hspace{1cm}w_1=-\frac{2p_1k}{2p_1p_2},\hspace{1cm}w_2=-\frac{2p_2k}{2p_1p_2},\hspace{1cm}x=\frac{(2p_1p_2) k^2}{(2p_1k)(2p_2 k)}.
\eeq

We are interested in the kinematic limit where the energies of the $m$ final state particles are almost zero, i.e., they are soft, and consequently we have $k\sim0$.
Furthermore, we are interested in the case where any loop momentum appearing in a virtual loop of our scattering process is low energetic as well. 
We refer to the resulting phase space and loop integrals as \emph{soft integrals}. Constructing such a soft integral follows the method of regions~\cite{Beneke:1997zp}, and details can be found in refs.~\cite{Anastasiou:2013srw,Anastasiou:2015yha}.

Above we only specified that the $m$ soft particles are in the final state. 
Depending on whether the remaining external particles are in the final or initial state, the soft integrals contribute to different kinds of scattering processes.
For example, the kinematic configuration where the momenta $p_1$ and $p_2$ in the initial state and $p_h$ in the final state corresponds to a partonic production process of the colorless state $h$ from hadron collisions at the LHC:
\beq
\text{ {\it Production}:}\hspace{1cm} p_1+p_2 \to p_h+k.
\eeq
If we consider $p_h$ and $p_1$ in the initial state and $p_2$ in the final state, we obtain a kinematic configuration corresponding to a semi-inclusive Deeply Inelastic Scattering (DIS) process:
\beq
\text{ {\it DIS}:}\hspace{1cm} p_1+p_h \to p_2+k.
\eeq
Finally, if only $p_h$ is in the initial state, this may be recognized as a doubly-resolved scattering configuration in $e^+ e^-$ annihilation:
\beq
\text{ {\it $e^+e^-$ Annihilation}:}\hspace{1cm} p_h\to p_1+p_2+k.
\eeq

In the remainder of this section, we will first discuss the final state phase space associated with the three different scattering configurations discussed above. 
Next, we will discuss the general structure of soft integrals, how to analytically continue it from one kinematic region to another, and finally we will define inclusive soft integrals.

\subsection{Final state phase space}

For different scattering processes we define the following phase space measures:
\begin{enumerate}
\item {\it Production}:
\beq
\df\Phi_{h+m}=(2\pi)^d \delta^d\left(p_1+p_2+p_h+\sum_{i=3}^{m+2}p_i\right)  \frac{\df^dp_h}{(2\pi)^d} (2\pi)\delta_+(p_h^2-m_h^2) \prod_{i=3}^{m+2} \frac{\df^dp_i}{(2\pi)^d} (2\pi)\delta_+(p_i^2).
\eeq
\item {\it DIS}:
\beq
\df\Phi_{1+m}=(2\pi)^d \delta^d\left(p_1+p_h+\sum_{i=2}^{m+2}p_i\right) \prod_{i=2}^{m+2} \frac{\df^dp_i}{(2\pi)^d} (2\pi)\delta_+(p_i^2).
\eeq
\item {\it $e^+e^-$ Annihilation}:
\beq
\df\Phi_{2+m}=(2\pi)^d \delta^d\left(p_h+\sum_{i=1}^{m+2}p_i\right) \prod_{i=1}^{m+2} \frac{\df^dp_i}{(2\pi)^d} (2\pi)\delta_+(p_i^2).
\eeq
\end{enumerate}
We work in dimensional regularization and denote the space time dimension by $d=4-2\epsilon$, where $\epsilon$ is the dimensional regulator.
Introducing the momentum $k$ and using the variables we defined in eq.~\eqref{eq:supervars}, we parametrise the phase space measures as follows:
\begin{enumerate}
\item {\it Production}:
\bea
\df\Phi_{h+m}&=&\frac{(s w_1 w_2)^{1-\epsilon}}{4(2\pi)^{3-2\epsilon}} (1-x)^{-\epsilon} \df\Phi_{m}(k)\, \df x\, \df w_1 \,\df w_2\, \df\Omega_{2-2\epsilon}\nonumber\\
&\times&\delta\left(\frac{m_h^2}{s}-(1-w_1-w_2+w_1 w_2 x)\right) \theta(x(1-x)) \theta(w_1) \theta(w_2).
\eea
\item {\it DIS}:
\bea
\df\Phi_{1+m}&=&\frac{(p_h^2w_1 w_2)^{1-\epsilon}}{4(2\pi)^{3-2\epsilon}} (1-w_1)^{-3+2\epsilon}(1-x)^{-\epsilon} \left(1-w_1-w_2+w_1 w_2 x\right)^{-1+\epsilon}\nonumber\\
&\times&\df\Phi_{m}(k)\, \df x\, \df w_1\, \df w_2\, \df\Omega_{2-2\epsilon}\, \theta(x(1-x)) \theta(-w_1) \theta(w_2)\theta(-p_h^2).
\eea
\item {\it $e^+e^-$ Annihilation}:
\bea
\df\Phi_{2+m}&=&
\frac{\pi ^{4-2 \epsilon } (p_h^2)^{1- \epsilon }(p_h^2 w_1w_2)^{1-\epsilon } (1-x)^{-\epsilon } }{(1-2 \epsilon )
   \Gamma (1-2 \epsilon )}(1-w_1-w_2+w_1 w_2 x)^{3 \epsilon -4} \nonumber\\
   &\times&\theta(x(1-x))\theta(-w_1)\theta(-w_2)\df w_1\, \df w_2\,\df x\,\df \Phi_{m}(k),\nonumber\\
\eea
\end{enumerate}
with
\beq
\df \Phi_{m}(k)=(2\pi)^d \delta^d\left(k-\sum_{i=3}^{m+3}p_i\right) \prod_{i=3}^{m+3} \frac{\df^dp_i}{(2\pi)^d} (2\pi)\delta_+(p_i^2).
\eeq
In the soft limit, where all final state partons associated with the momenta $p_3,\ldots,p_{m+2}$ become low energetic, we find
\beq
w_1\to 0,\hspace{1cm} w_2\to 0.
\eeq
In this limit, the differential phase space measures become proportional to each other. 
\bea
\lim_{w_{1,2}\to 0}\frac{\df \Phi_{h+m}}{\df w_1 \df w_2 \df x}&=&\frac{\df \Phi_{h+m}^{\text{soft}}}{\df w_1 \df w_2 \df x}=\frac{(p_h^2w_1 w_2)^{1-\epsilon}}{4(2\pi)^{3-2\epsilon}} (1-x)^{-\epsilon}\df\Phi_{m}(k)\, \df\Omega_{2-2\epsilon},\nonumber\\
\lim_{w_{1,2}\to 0}\frac{\df \Phi_{1+m}}{\df w_1 \df w_2 \df x}&=&\lim_{w_{1,2}\to 0}\frac{\df \Phi_{h+m}}{\df w_1 \df w_2 \df x},\nonumber\\
\lim_{w_{1,2}\to 0}\frac{\df \Phi_{2+m}}{\df w_1 \df w_2 \df x}&=&\frac{(4 \pi )^{\epsilon -2} (p_h^2)^{1- \epsilon } \Gamma (1-\epsilon )}{(1-2 \epsilon ) \Gamma (1-2 \epsilon )} \lim_{w_{1,2}\to 0}\frac{\df \Phi_{h+m}}{\df w_1 \df w_2 \df x}.
\eea
Above, we implicitly set Kronecker delta constraints of external variables to unity.

\subsection{General structure of soft integrals}
We define a differential soft Feynman integral including $L$ loops and $m$ soft final state particles by
\beq
\label{eq:generalsoftint}
I^\text{D}(s,w_1,w_2,x)=c_{L+m} \int \frac{\df \Phi_{h+m}^{\text{soft}}}{\df w_1 \df w_2 \df x} \prod_{i=0}^L \frac{\df^d p_{2+m+i}}{(2\pi)^d} \mathcal{I}^{\text{D}}(p_h,p_1,\dots,p_{2+L+m},\epsilon).
\eeq
We define the constant
\beq
c_L=(4\pi)^{-L\epsilon} e^{L \epsilon \gamma_E}.
\eeq
The integrand $\mathcal{I}^\text{D}$ is a ratio of polynomials in Lorentz invariant scalar products of the external momenta and loop momenta as well as the dimensional regulator. 
The superscript label `D' indicates that we refer to this integrals as {\it differential} soft integrals. 
In contrast, we define {\it inclusive} soft integrals with superscript `I' as 
\beq\begin{split}
\label{eq:incdiff}
I^{\text{I}}(s,\bar z)&\,=c_{L+m}\int\df \Phi_{h+m}^{\text{soft}} \,\prod_{i=0}^L \frac{\df^d p_{2+m+i}}{(2\pi)^d} \,\mathcal{I}^{\text{I}}(p_h,p_1,\dots,p_{2+L+m},\epsilon)\\
&\,=\int_0^1 \df x \int_0^\infty \df w_1\, \df w_2\, \delta(\bar z-w_1-w_2)\,I^\text{D}(s,w_1,w_2,x).
\end{split}\eeq
For example, the inclusive soft phase space volume is given by
\bea
\label{eq:softvol}
\int \df\Phi_{h+m}^{\text{soft}}&=&\int_0^1 \df x \int_0^\infty \df w_1\, \df w_2\, \delta(\bar z-w_1-w_2)\lim_{w_{1,2}\to 0}\frac{\df \Phi_{h+m}}{\df w_1 \df w_2 \df x}\nonumber\\
&=&\frac{(4 \pi)^{1-2\epsilon m} e^{\gamma_E \epsilon m} (p_h^2)^{-1+m-m\epsilon} \bar{z}^{-1+2 m(1-\epsilon )} \Gamma (1-\epsilon )^m}{2\Gamma (2 m (1-\epsilon ))}.
\eea

Properly chosen integrands of soft integrals are characterized by a rescaling symmetry. 
\bea
 \mathcal{I}^{\text{D}}(p_h, \lambda p_1,p_2,\dots,p_{2+L+m},\epsilon)&=&\lambda^{\alpha_1} \mathcal{I}^{\text{D}}(p_h,p_1,p_2,\dots,p_{2+L+m},\epsilon),\nonumber\\
 \label{eq:rescaling}
 \mathcal{I}^{\text{D}}(p_h,  p_1,\lambda p_2,\dots,p_{2+L+m},\epsilon)&=&\lambda^{\alpha_2} \mathcal{I}^{\text{D}}(p_h,p_1,p_2,\dots,p_{2+L+m},\epsilon),
\eea
for integer exponents $\alpha_1$ and $\alpha_2$ that can be determined easily from the specific integrands. 
This is most easily illustrated looking at an example. 
For the following integrand of a phase space integral with two additional partons in the final state, we find the exponents $\alpha_1=\alpha_2=-1$:
\beq
 \mathcal{I}^{\text{D}}_{\text{example}}(p_h,  p_1,p_2,p_3,p_4,\epsilon)=\frac{1}{(2p_1 p_3)(2p_2p_4)}.
\eeq
From the fact $s$ is the only Lorentz invariant variable in our set of variables of eq.~\eqref{eq:supervars}, we conclude that any differential soft integral takes the form
\beq
\label{eq:softstructure}
I^\text{D}(s,w_1,w_2,x)=s^{\Lambda-(m+L)\epsilon} w_1^{\delta_1-(m+L)\epsilon}w_2^{\delta_2-(m+L)\epsilon} f(x,\epsilon).
\eeq
Above, the integer mass dimension $\Lambda$ and the integer exponents $\delta_1$ and $\delta_2$ can be determined from the integrand using the rescaling symmetry in conjunction with the dependence of the loop and phase space measure on the variable $s$, $w_1$ and $w_2$. 
The integers $m$ and $L$ are the number of final state partons that were integrated out and the number of loops. 
The variable $x$ is invariant under a rescaling of the momenta $p_1$ and $p_2$, and consequently our soft differential integrals  have a non-trivial functional dependence on $x$ in the form of a function $f(x,\epsilon)$.
Inclusive soft master integrals then take the form
\beq
\label{eq:genericsoft}
I^\text{I}(s,\bar z)=s^{\Lambda-(m+L)\epsilon} \bar z^{\delta_1+\delta_2+1-2(m+L)\epsilon} \tilde f(\epsilon),
\eeq
where $\bar z$ is introduced via eq.~\eqref{eq:incdiff} and $\tilde f(\epsilon)$ is a function of the dimensional regulator.

Having identified the structure of differential soft integrals, we can now discuss what happens when crossing from production to DIS or $e^+ e^-$ kinematics.
First, we note that the variable $x$ is by definition (eq.~\eqref{eq:supervars}) invariant under crossing the partons with momenta $p_1$ or $p_2$ from initial to final state or vice versa. 
We can collect the dependence of soft integrals on the remaining variables $s$, $w_1$ and $w_2$ using eq.~\eqref{eq:softstructure} into one prefactor.
\beq
I^\text{D}\sim \left(s w_1 w_2 \right)^{-(m+L)\epsilon}=\left(\frac{ (2p_1 k)(2p_2 k)}{(2p_1 p_2)} \right)^{-(m+L)\epsilon}.
\eeq
From the above equation we easily see that this factor is also invariant under crossing $p_1$ or $p_2$ from the initial to the final state or vice versa. 
Consequently, differential soft integrals are identical for production, DIS or $e^+ e^-$ scattering kinematics, up to an overall sign that can be determined from the integer powers $\Lambda$, $\delta_1$ and $\delta_2$.
\section{Computing soft master integrals}
\label{sec:SoftIntegrals}
We begin this section by outlining our method to compute soft integrals. 
After that, we show our explicit results for phase space integrals with two additional final state partons, ($m=2$, $L=0$ in eq.~\eqref{eq:generalsoftint}). 
We refer to these integrals as double real (RR) phase space integrals.
We then briefly discuss our results for soft integrals with three additional partons in the final state (tripple real; RRR) and soft integrals with two additional partons in the final state and one loop integral (double-real virtual; RRV).

\subsection{Method}
Soft Feynman integrals can be related to each other via the framework of reverse unitarity~\cite{Anastasiou2002,Anastasiou2003,Anastasiou:2002qz,Anastasiou:2003yy,Anastasiou2004a} and IBP identities~\cite{Tkachov1981,Chetyrkin1981,Laporta:2001dd}.
We construct a basis of master integrals for soft integrals involving a certain number of loop and phase space integrals. 
We then use methods developed in refs.~\cite{Lee:2016bib,Henn:2020lye} to construct a basis of so-called canonical master integrals.
We construct such a basis for both differential and inclusive soft integrals. 
We use differential equations~\cite{Kotikov:1990kg,Kotikov:1991hm,Kotikov:1991pm,Gehrmann:1999as,Henn:2013pwa} for differential soft master integrals to compute the functional dependence of these integrals on the variable $x$. Next, we use eq.~\eqref{eq:incdiff} to relate the differential and inclusive soft master integrals to each other. 
Since in many cases the inclusive soft master integrals can be evaluated directly, without first computing their differential analogues, 
we can express the boundary conditions required for the solution of the differential equations for the differential soft master integrals in terms of the inclusive soft master integrals. 
This relation between inclusive and differential master integrals also constrains some of the inclusive master integrals.
We elaborate on this below in section~\ref{sec:relations} using an example.
Additional consistency conditions that can be determined from the system of differential equations and from relations of the differential soft master integrals to systems of differential equations appearing in the computation of ref.~\cite{Ebert:2020yqt,Ebert:2020unb}, and this gives additional constraints on the inclusive soft master integrals. 
Ultimately, we determine the remaining inclusive soft master integrals using direct integration techniques developed in refs.~\cite{Anastasiou:2013srw,Anastasiou:2015yha}. 
We express the inclusive soft master integrals as a Laurent series in the differential regulator with rational numbers and the multiple $\zeta$ values as coefficients (and they depend on $p_h^2$ and $\bar z$ as shown in eq.~\eqref{eq:genericsoft}). 
The resulting differential master integrals depend on the variables $p_h^2$, $w_1$ and $w_2$, as illustrated in eq.~\eqref{eq:softstructure}, and the function $f(x)$ is given by a Laurent series in the dimensional regulator and harmonic polylogarithms~\cite{Remiddi:1999ew} with argument $x$ and multiple $\zeta$ values as coefficients.
As we choose canonical integrals as our basis integrals, the master integrals have uniform transcendental weight. 
In particular, we compute in this article all soft master integrals up to transcendental weight eight, or equivalently up to $\mathcal{O}(\epsilon^8)$ in the Laurent expansion.

\subsection{RR soft integrals with two final state partons}
As an example, we discuss explicitly in this section the soft master integrals for pure phase space integrals with two partons in the final state that are integrated out (RR).
The corresponding inclusive soft master integrals were presented already to all orders in the dimensional regulator in ref.~\cite{Anastasiou:2012kq}.
\subsubsection{Differential RR soft master integrals}
We define three differential RR soft master integrals by
\beq
I_{i}^{\text{D-RR}}=c_2 \int \frac{\df \Phi_{h+2}^{\text{soft}}}{\df w_1 \df w_2 \df x} \mathcal{I}_{i}^{\text{D-RR}}.
\eeq
The corresponding integrands are given by
\bea
\label{eq:diffdef}
\mathcal{I}_{1}^{\text{D-RR}}&=&-\frac{s_{12}^2 \epsilon  (s_{13}+s_{14})}{s_{14}},\nonumber\\
\mathcal{I}_{2}^{\text{D-RR}}&=&\frac{s_{12}^2 \epsilon  \left(s_{12}^2 s_{34}-s_{14} s_{23}-s_{14} s_{24}\right)}{s_{14} s_{24}},\\
\mathcal{I}_{3}^{\text{D-RR}}&=&\frac{s_{12}^2 \epsilon  \left(s_{12}^2 s_{34}-s_{12} s_{13} s_{23}-s_{12} s_{13} s_{24}-s_{12} s_{14} s_{23}-s_{12} s_{14} s_{24}+s_{13} s_{23}+s_{13} s_{24}\right)}{s_{13} s_{24}}.\nonumber
\eea
Here, we used the notation
\beq
s_{ij}=(p_i+p_j)^2, \hspace{1cm}s_{ii}=p_i^2.
\eeq
The integrated results in terms of our chosen variables of eq.~\eqref{eq:supervars} are given by
\bea
I_{1}^{\text{D-RR}}&=&\frac{e^{2 \gamma  \epsilon } s_{12}^{-2 \epsilon } w_1^{-2 \epsilon } w_2^{-2 \epsilon } (1-x)^{-\epsilon } x^{-\epsilon }}{128 \pi ^3 \Gamma (1-2 \epsilon )}.\\
I_{2}^{\text{D-RR}}&=&I_{1}^{\text{D-RR}} (1-2 x \, _2F_1(1,1;1-\epsilon ;1-x)).\nonumber\\
I_{3}^{\text{D-RR}}&=&I_{1}^{\text{D-RR}}(-2 x \, _2F_1(1,1;\epsilon +1;1-x)+2 x^{\epsilon } (1-x)^{-\epsilon } \Gamma (1-\epsilon ) \Gamma (\epsilon +1)+1).\nonumber
\eea
These results are valid to all orders in the dimensional regulator, and ${}_2 F_1$ is the Gauss hypergeometric function,
\beq
{}_2F_1(a,b;c;x) = \sum_{n=0}^\infty\frac{(a)_n(b)_n}{(c)_n}\frac{x^n}{n!}\,.
\eeq
This function can be easily expanded in terms of a Laurent series in the dimensional regulator using the results of ref.~\cite{Huber:2005yg}.

\subsubsection{Inclusive RR soft master integrals}
To express a basis of double real inclusive soft master integrals we require two different master integrals.
\beq
I_{i}^{\text{I-RR}}=c_2 \int \df\Phi_{h+2}^{\text{soft}} \mathcal{I}_{i}^{\text{I-RR}}.
\eeq
The integrands are given by
\bea
 \mathcal{I}_{1}^{\text{I-RR}}&=&\epsilon^3 \frac{s_{12}}{s_{14} s_{34}},\nonumber\\
 \mathcal{I}_{1}^{\text{I-RR}}&=&\epsilon^3\bar z \frac{s_{12}^2}{s_{13} s_{24}s_{34}}.
\eea
The integrals are given by
\bea
I_{1}^{\text{I-RR}}&=&(1-2 \epsilon ) (3-4 \epsilon ) (1-4 \epsilon )\Phi_{h+2}^{\text{soft}}=\frac{e^{-2 \gamma  \epsilon } \bar z ^{-4\epsilon }s_{12}^{-2 \epsilon} \Gamma (1-\epsilon )^2}{256 \pi ^3  \Gamma (1-4 \epsilon )},\nonumber\\
I_{2}^{\text{I-RR}}&=&-\frac{3 e^{-2 \gamma  \epsilon } \Gamma (1-2 \epsilon )^2 \Gamma (1-\epsilon )}{128 \pi ^3  \Gamma (1-4 \epsilon ) \Gamma (1-3 \epsilon )}  \bar z ^{-4\epsilon }s_{12}^{-2 \epsilon} \, _3F_2(-\epsilon ,-\epsilon ,-\epsilon ;1-\epsilon ,-3 \epsilon ;1).
\label{eq:X18sol}
\eea

\subsubsection{Relating inclusive and differential RR soft master integrals}
\label{sec:relations}
The fact that we compute a basis of master integrals simulatneously for the differential and inclusive cases can be very helpful. 
If we integrate the differential integrals over the inclusive soft phase space measure using IBP identities, we find that the result is related to the phase space volume.
\bea
c_2 \int \df\Phi_{h+2}^{\text{soft}} \mathcal{I}_{1}^{\text{D-RR}}&=&(1-2\epsilon)\Phi_{h+2}^{\text{soft}}, \nonumber\\
c_2 \int \df\Phi_{h+2}^{\text{soft}} \mathcal{I}_{2}^{\text{D-RR}}&=&-\Phi_{h+2}^{\text{soft}}, \nonumber\\
c_2 \int \df\Phi_{h+2}^{\text{soft}} \mathcal{I}_{3}^{\text{D-RR}}&=&\Phi_{h+2}^{\text{soft}}. 
\eea
Since we are computing the differential master integrals using the method of differential equations, we can fix the boundary conditions of the differential equations by performing the inclusive integration over all differential variables and demanding that the above equations are true.
Conversely, we may integrate the integrand of an inclusive soft master integral over the differential soft measure.
\beq
c_2 \int \frac{\df \Phi_{h+2}^{\text{soft}}}{\df w_1 \df w_2 \df x} \mathcal{I}_{2}^{\text{I-RR}}=-\frac{s_{12}^2 \epsilon^2}{x(1-x)w_1 w_2}\left[I_{1}^{\text{D-RR}}+I_{3}^{\text{D-RR}}\right]
\eeq
Subsequently integrating over the remaining variables $w_1$, $w_2$ and $x$ we can determine the value of the inclusive master integral $I_{2}^{\text{I-RR}}$ and indeed find the solution of eq.~\eqref{eq:X18sol}.
\beq
I_{2}^{\text{I-RR}}=c_2\int \df x\df w_1 \df w_2 \delta(1-w_1-w_2) \int \frac{\df \Phi_{h+2}^{\text{soft}}}{\df w_1 \df w_2 \df x} \mathcal{I}_{2}^{\text{D-RR}}.
\eeq
In this fashion, we determined all boundary conditions and inclusive RR soft master integrals by only computing the inclusive soft phase space volume (eq.~\eqref{eq:softvol}) explicitly. 
We observe in more complicated cases than the RR soft master integrals that additional boundary conditions need to be computed by other means.

\subsection{RRR and RRV soft master integrals}
One of the main results of this article is the computation of differential and inclusive soft master integrals through transcendental weight eight for RRV and RRR scattering configurations. 
To determine these integrals, we follow the steps outlined above, and we present our results in terms of computer readable files together with the arXiv submission of this article. In particular, we compute 24 differential and 14 inclusive RRV soft master integrals.
\bea
I_{i}^{\text{D-RRV}}&=&c_3 \int \frac{\df \Phi_{h+2}^{\text{soft}}}{\df w_1 \df w_2 \df x} \int \frac{\df^dp_6}{(2\pi)^d} \,\mathcal{I}_{i}^{\text{D-RRV}},\\
I_{i}^{\text{I-RRV}}&=&c_3 \int\df \Phi_{h+2}^{\text{soft}}  \int \frac{\df^dp_6}{(2\pi)^d}\,\mathcal{I}_{i}^{\text{I-RRV}}.
\eea
Furthermore, we compute 61 differential and 13 inclusive soft RRR master integrals.
\bea
I_{i}^{\text{D-RRR}}&=&c_3 \int \frac{\df \Phi_{h+3}^{\text{soft}}}{\df w_1 \df w_2 \df x}\, \mathcal{I}_{i}^{\text{D-RRR}},\\
I_{i}^{\text{I-RRR}}&=&c_3 \int \df\Phi_{h+3}^{\text{soft}}\, \mathcal{I}_{i}^{\text{I-RRR}}.
\eea
Many of our integrals were computed already through transcendental weight six for the purpose of refs.~\cite{Anastasiou:2013srw,Anastasiou:2015yha,Dulat:2017prg,Ebert:2020yqt,Ebert:2020unb}, and we find agreement with these past results. 
In order to ensure the correctness of our soft integrals, we furthermore explicitly derive numerical results for them through all calculated orders in $\epsilon$ using Mellin-Barnes (MB) techniques. Indeed, the properties of the integrands under rescaling in eq.~\eqref{eq:rescaling} implies that for RRR we can easily integrate out the energies of the soft particles in terms of $\Gamma$ functions. The remaining angular integrals can be perform in closed form as a MB integral~\cite{Somogyi:2011ir}. Following this strategy, we can easily obtain an MB representation for all RRR soft integrals~\cite{Anastasiou:2013srw}, which can be evaluated numerically as a Laurent series in the dimensional regulator using standard techniques~\cite{Czakon:2005rk,Smirnov:2009up}. For the RRV integrals, we cannot immediately perform the integration over the energies in terms of $\Gamma$ functions, because they are entangled with the loop integration. We can, however, easily introduce additional MB integrations for the (soft regions of the) one-loop integrals involved~\cite{Anastasiou:2015yha}, and and then evaluate numerically the MB representations for the combined phase space and loop integrations.

\section{Threshold limit of  production cross sections}
\label{sec:thresholdxs}
In this section we discuss the LHC cross sections for the production of a virtual photon or a Higgs boson in gluon fusion in the infinite top quark mass limit:
\beq
\label{eq:hadronicxs}
\sigma_B=\tau \hat \sigma_0^BC^2_B \sum_{ij} f_i(\tau) \circ_\tau \eta_{ij}^B(\tau) \circ_\tau f_j(\tau) ,\hspace{1cm} B\in\{H,\gamma^*\}.
\eeq
In the above equation the $f_i$ are parton distribution function, $\hat \sigma_0^B$ represents the partonic Born cross section and we define the ratio $\tau=Q^2/ S$, where $Q$ is the virtuality of the produced boson\footnote{In case of the Higgs boson production cross section, $Q$ is identical to the Higgs boson mass.} and $S$ is the hadronic centre of mass energy. The PDFs are convoluted with the partonic coefficient functions, see appendix~\ref{sec:convolute} for details.
The partonic coefficient functions are given by
\beq
\label{eq:PCF}
\eta^B_{ij}(z)=\frac{\mathcal{N}_{ij}}{2Q^2 \hat \sigma_0^B}  \sum_{m=0 }^\infty \int \df \Phi_{h+m}\mathcal{M}_{ij\to B+m}.
\eeq
The initial state dependent normalisation factor $\mathcal{N}_{ij}$ is given by
\beq
\mathcal{N}_{gg}=\frac{1}{4(n_c^2-1)^2(1-\epsilon)^2},\hspace{1cm}
\mathcal{N}_{gq}=\frac{1}{4(n_c^2-1)n_c(1-\epsilon)},\hspace{1cm}
\mathcal{N}_{q\bar q}=\frac{1}{4n_c^2},
\eeq
where $g$, $q$ and $\bar q$ represent a gluon, quark and anti-quark respectively, and $n_c$ denotes the number of fundamental SU($n_c$) colors. The factor $C_B$ is equal to one for the production cross section of a virtual photon and equal to the Wilson coefficient~\cite{Chetyrkin:1997un,Schroder:2005hy,Chetyrkin:2005ia,Kramer:1996iq} for the infinite top quark mass effective field theory~\cite{Inami1983,Shifman1978,Spiridonov:1988md,Wilczek1977}. $\mathcal{M}_{ij\to B+m}$ represents the color and spinor summed interference of  scattering amplitudes describing the production of the desired boson $B$ and $m$ final state partons in the collision of initial state partons $i$ and $j$. Expanding $\mathcal{M}_{ij\to B+m}$ in the strong coupling constant $\alpha_S$ gives the perturbative coefficient functions in QCD perturbation theory.
\beq
\eta_{ij}^B(z)=\sum_{i=0}^\infty a_S^i \eta_{ij}^{B,\,(i)}(z),\hspace{1cm} a_S=\frac{\alpha_S}{\pi} (4\pi)^{\epsilon} e^{-\epsilon \gamma_E}.
\eeq
In the definition of $a_S$ we include for convenience already a factor $(4\pi)^{\epsilon} e^{-\epsilon \gamma_E}$, anticipating later $\overline{\text{MS}}$ renormalisation.
At Born level we find
\beq
\eta_{gg}^{H,\,(0)}(z)=\frac{1}{1-\epsilon} \delta(\bar z),\hspace{1cm}
\eta_{q \bar q}^{\gamma^*,\,(0)}(z)=(1-\epsilon) \delta(\bar z), \hspace{1cm} \bar z=1-z.
\eeq
For more details see refs.~\cite{Mistlberger:2018etf,Duhr:2020seh}.

In this article, we are interested in the threshold limit of the partonic coefficient function. 
This limit is characterized by the kinematic condition that all the radiation produced along side the produced boson is very low energetic, and we consider the limit $\bar z \to 0$.
In this limit the partonic coefficient function factorizes as follows.
\beq
\lim\limits_{\bar z\to 0}\eta_{ij}^B=H_{ij}^B \times S^{R_{ij}}_{\text{thr.}}(z).
\eeq
Above, $H_{ij}^B$ is the process dependent hard function and $S^{R_{ij}}_{\text{thr.}}$ is the so-called threshold soft function that only depends on the color representation of the initial state partons $R_{ij}$. 
The hard function was computed for Higgs boson and photon production through three loops in refs.~\cite{Gehrmann:2005pd,Baikov:2009bg,Gehrmann:2010ue,Gehrmann:2010tu} and at fourth loop order in refs.~\cite{Henn:2019swt,vonManteuffel:2020vjv,Agarwal:2021zft,Lee:2021uqq,Lee:2022nhh}.
Similarly, $S^{R_{ij}}_{\text{thr.}}$ was computed through N$^3$LO in QCD perturbation theory in refs.~\cite{Anastasiou:2014vaa,Li:2014afw,Ahmed:2014cla} and we discuss partial fourth loop-order results below.

The threshold soft function can be computed by considering the strict soft limit of the partonic coefficient function of one of Higgs boson or photon production. 
\bea
\label{eq:xsSFconnection}
S^{\text{adjoint}}_{\text{thr.}} (z)&=&(1-\epsilon) \lim\limits_{\text{strict soft}} \eta_{ij}^H (z),\nonumber\\
S^{\text{fundamental}}_{\text{thr.}} (z)&=&\frac{1}{(1-\epsilon)} \lim\limits_{\text{strict soft}} \eta_{ij}^{\gamma^*} (z).
\eea
The strict soft limit is defined by taking all final state parton momenta to be very low energetic and all loop momenta to be uniformly low energetic as well. 
It is now easy to see that the partonic cross section in this limit can be expressed in terms of soft master integrals as introduced in previous sections. 
In practice, this is achieved by following the method of regions~\cite{Beneke:1997zp} and using techiques introduced in refs.~\cite{Anastasiou:2013srw,Anastasiou:2015yha}.
To compute the threshold soft function contribution at $n^{\text{th}}$ perturbative order, matrix elements with up to $n$ additional soft partons in the final state must be included. 
Contributions with one additional parton can be extracted from the computation of the one emission current at one and two loop order~\cite{Catani:2000pi,Duhr:2013msa,Li:2013lsa,Dixon:2019lnw}. 
The integrand for two and three additional partons required for computations up to N$^3$LO in QCD perturbation theory was determined for the purposes of refs.~\cite{Anastasiou:2013srw,Anastasiou:2014vaa,Anastasiou:2015yha,Ebert:2020yqt,Ebert:2020unb,Dulat:2017prg}, and we build on these results here. 
In particular, we use our newly computed soft master integrals to compute $S^{R_{ij}}_{\text{thr.}}(z)$ at N$^3$LO in QCD perturbation theory. 
This result was previously obtained in refs.~\cite{Anastasiou:2014vaa,Li:2014afw,Ahmed:2014cla} through finite order in the dimensional regulator $\eps$ at N$^3$LO, and we find agreement. 
We extend this results here to include two additional orders in the Laurent expansion in the dimensional regulator, which will serve as a key ingredient for a future computation of the Higgs boson and Drell-Yan production cross section at N$^4$LO.
In particular, we compute results for the bare partonic cross section $\eta_{ij}^B(z)$ in the threshold limit to $\mathcal{O}(\epsilon^{8-2n})$ at N$^n$LO in QCD perturbation theory for $n\in\{0,1,2,3\}$. 
We attach our results in electronically readable form alongside the arXiv submission of this article.

\section{Threshold factorization and soft anomalous dimension}
\label{sec:fac}
\subsection{Threshold factorisation}
The inclusive cross section for the production of a colorless final state factorizes in the limit where the hadronic center-of-mass energy becomes similar to the invariant mass of the colorless system, i.e., $\tau\to 1$.
This was realized in refs.~\cite{Sterman:1986aj,Catani:2003zt,Catani:1989ne,Catani:1990rp} for QCD and derived in the language of soft-collinear effective theory (SCET)~\cite{Bauer:2000ew,Bauer:2000yr,Bauer:2001ct,Bauer:2001yt,Bauer:2002nz} in refs.~\cite{Ahrens:2009cxz,Ahrens:2008qu,Ahrens:2010rs}.
Mathematically, we may write eq.~\eqref{eq:hadronicxs} in this limit as\footnote{For simplicity, we set here $C_B=1$ as it can easily be absorbed in the hard function.}
\beq
\label{eq:thresholdxs}
\sigma_B= \hat \sigma_0^B \sum_{ij} H_{ij}^B f_i^{\text{th}}(\tau) \otimes_\tau S_{\text{thr.}}^r(\tau) \otimes_\tau f_j^{\text{th}}(\tau) +\mathcal{O}(1-\tau).
\eeq
Above, the product $\otimes_{\tau}$ is defined in the appendix in eq.~\eqref{eq:softconv} and $r=R_{ij}$ denotes the color representation of the initial state partons.
 The hard function $ H_{ij}^B(\mu^2)$ is the squared Wilson coefficient of the leading power hard scattering operator that couples the color singlet to the partons $i$ and $j$. It is related to the form factor of such an operator. 
 
 \begin{figure}[h!]
 \centering
 \includegraphics[width=0.29\textwidth]{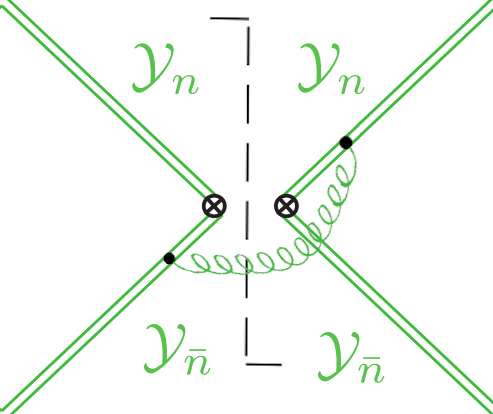}
 \caption{Cut diagram for the one real emission correction to the threshold soft function $\Sth$ for color singlet production. 
 Doubled lines represent Wilson lines that trace the path of the initial state partons involved in the hard scattering process. 
 The wiggled line represents a radiated gluon which crosses the final state phase space cut indicated by the dashed line.
}
 
 \label{fig:SF}
\end{figure}
 We can define the soft function as a squared matrix element of soft Wilson lines \cite{Korchemsky:1993uz,Belitsky:1998tc} involving a measurement over a complete set of soft states $\ket{X_s}$,
\beq \label{eq:Sthdef}
	S_{\text{thr.}}^r(z)=\frac{1}{C_r} \sum_{X_s} \tr \bra{0} Y^{\dagger\,r}_\bn(0) Y^r_n(0)\delta(\hat{E}^2 - {Q^2(1-z)}) \ket{X_s}\bra{X_s} Y^{\dagger\,r}_n(0) Y^r_\bn(0) \ket{0}   \,,
\eeq
where the operator $\hat{E}$ picks up the total energy of the real emissions, the Wilson lines are taken in the representation $r$ of the scattering partons and along their light-like direction $n,\bn$ and $C_r$ is the quadratic Casimir of the gauge group for the representation $r$. Note that for the rest of this section, we will drop the label $r$ for the representation, but the following discussion straightforwardly applies to both the case of adjoint and fundamental representation.
Explicitly the Wilson lines take the form
\beq
 Y_n (x) = \mathbf{P} \exp{\left[ig \int_{-\infty}^0 \df t\, n \cdot A_s(x +t n) \right]}\,.
\eeq 
 The contributions to $S_{\text{thr.}}^r(z)$ can be represented, order by order in perturbation theory, in terms of cut diagrams and can be calculated in terms of eikonal Feynman rules, see for example \fig{SF}.

 The quantities in eq~\eqref{eq:thresholdxs} are bare quantities that are individually ultraviolet and infrared divergent, and the divergences manifest themselves as poles in the dimensional regulator $\epsilon$.
 We implement the renormalisation of the strong coupling constant via the operator ${\bf Z}_{\alpha_S}$ (see appendix~\ref{sec:uvren}), which expresses the bare strong coupling constant in terms of its renormalised counterpart.
We absorb infrared singularities of the hard function into $Z_H(\mu^2)$.
 The threshold PDFs $f^{\text{th}}_i(\tau)$ absorb collinear initial state singularities via a standard mass factorization counter term (see appendix~\ref{sec:convolute} for details).
 \bea
 \label{eq:renormalise}
 H_{ij}^{B}(\mu^2)&=&Z_H(\mu^2){\bf Z}_{\alpha_S} H_{ij}^{B},\nonumber\\
f^{\text{th}}_{i}(\tau,\mu^2)&=&\Gamma^r(\tau)\otimes_\tau f_{i}^{\text{th}}(\tau),\nonumber\\
S_{\text{thr.}}^r(\tau,\mu^2)&=&Z_H(\mu^2)^{-1} \Gamma^r(\tau)^{-1}\otimes_\tau {\bf{Z}}_{\alpha_S}S_{\text{thr,}}^r(\tau)\otimes_\tau \Gamma^r(\tau)^{-1}.
 \eea
 Note, that we indicate renormalized and finite objects by explicitly indicating their dependence on the scale $\mu^2$. 
 
 In the context of SCET, the divergences appearing in the bare soft function can be interpreted as UV divergences in the effective theory. Therefore, we can absorbe these divergences in an $\MSbar$ counterterm $Z_S(z,\mu^2)$ for this operator and obtain a renormalized soft function.  
 Extracting from eq.~\eqref{eq:renormalise} we find
 \bea
 \label{eq:SFfiniteDef}
 \Sth(z,\mu^2)&=&Z_S(z,\mu^2)\otimes_z {\bf Z}_{\alpha_S}\Sth(z),\nonumber\\
Z_S(z,\mu^2)\ &=&Z_H(\mu^2)^{-1} \Gamma^r(z)^{-1}\otimes_z  \Gamma^r(z)^{-1}.
 \eea
 
 The hard function obeys the renormalisation group equation (RGE)
\bea
\label{eq:HRGE}
	\mu^2 \frac{\df }{\df \mu^2} H_{ij}^B(\mu^2)& =& \gamma_H^r (\alpha_S(\mu^2),\mu^2) H_{ij}^B(\mu^2),\nonumber\\
	\gamma_H^r (\alpha_S(\mu^2),\mu^2) & =& \GammaC^r(\alpha_S(\mu))\ln \frac{Q^2}{\mu^2} + \frac{1}{2}\gamma_H(\alpha_S(\mu^2)).
\eea
Above $\Gamma_\text{cusp}^r(\alpha_S(\mu^2))$ is the cusp anomalous dimension~\cite{Korchemsky:1993uz}.
Furthermore, we distinguish $\gamma_H^r(\alpha_S(\mu^2))$, the non-cusp part of the anomalous dimension, from the entire anomalous dimension $ \gamma_H^r (\alpha_S(\mu^2),\mu^2)$ by the number of arguments.
  The renormalisation group equation for the threshold PDFs is given by the DGLAP evolution equation~\cite{Altarelli:1977zs,Dokshitzer:1977sg,Gribov:1972ri} in the limit of $z\to 1$.
  Explicitly, the anomalous dimension of the threshold PDFs is the limit of $z\to 1$ of the Altarelli-Parisi splitting functions:
\bea
\nonumber \gamma_f^{\text{fundamental}}(z,\alpha_S(\mu^2))&=\lim_{z\to 1} P_{qq}(z,\alpha_S(\mu^2)),\\
\gamma_f^{\text{adjoint}}(z,\alpha_S(\mu^2))&=\lim_{z\to 1} P_{gg}(z,\alpha_S(\mu^2)).
\eea
With this we find
 \bea
	\mu^2 \frac{\df }{\df \mu^2} f^{\text{th}}_{i}(z,\mu^2)&=&\gamma_f^r(z,\alpha_S(\mu^2))\otimes_z  f^{\text{th}}_{i}(z,\mu^2),\nonumber\\
	\gamma_f^r(z,\alpha_S(\mu^2))&=&\Gamma_{\text{cusp}}^r(\alpha_S(\mu^2))  \left[\frac{1}{1-z}\right]_++\frac{1}{2} \gamma_f^r(\alpha_S(\mu^2)) \delta(1-z).
 \eea
Above, the plus distribution is defined via its action on a test function as
\beq
\int_0^1 dz \left[\frac{1}{1-z}\right]_+ \phi(z)=
\int_0^1 dz \left(\frac{1}{1-z}\right) (\phi(z)-\phi(1)).
\eeq

 Since the hadronic cross section is independent of the scale $\mu^2$, the threshold soft function also satisfies an RGE that can be derived by consistency:
 \bea 
\nonumber	\mu^2 \frac{\df }{\df \mu^2}  \Sth(z,\mu^2) &=&\gamma_{\text{thr.}}^r(z,\alpha_S(\mu^2))\otimes_z  \Sth(z,\mu^2) ,\\
	\gamma_{\text{thr.}}^r(z,\alpha_S(\mu^2))&=&-\Gamma_{\text{cusp}}^r(\alpha_S(\mu^2)) \left(2 \left[\frac{1}{1-z}\right]_++\delta(1-z) \log\frac{Q^2}{\mu^2}\right)\\
	&&+\frac{1}{2} \gamma_{\text{thr.}}^r(\alpha_S(\mu^2)) \delta(1-z).\nonumber
 \eea
In order for the hadronic cross section in eq.~\eqref{eq:thresholdxs} to be independent of the scale, the following equation has to be satisfied.
\beq
\label{eq:gammathr}
\gamma_{\text{thr.}}^r(\alpha_S(\mu^2))=-2 \gamma_f^r(\alpha_S(\mu^2)) -\gamma_H^r(\alpha_S(\mu^2)).
\eeq


\subsection{Results for the threshold anomalous dimension through N$^4$LO}
All but one ingredient to compute the threshold limit of the partonic coefficient function for DY and Higgs boson production at N$^4$LO are currently available. 
The one missing ingredient is the coefficient of $\delta(\bar z)$ of the threshold soft function at N$^4$LO. 
The hard function can be extracted from the computation of the purely virtual matrix elements computed through four loops in QCD in refs.~\cite{Gehrmann:2010ue,vonManteuffel:2020vjv,Lee:2021uqq,Agarwal:2021zft,Lee:2022nhh,Baikov:2009bg,Gehrmann:2010tu,Gehrmann:2005pd}.
The ultraviolet renormalisation counterterms are given in terms of the QCD beta function~\cite{Baikov:2016tgj,Herzog:2017ohr,vanRitbergen:1997va,Czakon:2004bu,Tarasov:1980au,Larin:1993tp}. 
The cusp anomalous dimension is known through fourth loop order~\cite{vonManteuffel:2020vjv,Henn:2019swt}.
The mass factorization counterterm at threshold can be determined from the Altarelli-Parisi splitting functions, which have been determined through second~\cite{Moch:2015usa,Ablinger:2014nga,Moch:2004pa,Moch:2014sna,Vogt:2004mw,Ablinger:2017tan,Blumlein:2021enk} and third~\cite{Vogt:2018miu,Das:2020adl,Moch:2017uml,Moch:2021qrk,Moch:2018wjh,Das:2019btv} non-trivial order. 
The last ingredient is not yet available in fully analytic form, and some constants have been determined only numerically, based on an extraction from a computation of several moments of the full splitting functions and a leading color computation. 
The consequence is that the splitting functions are afflicted by an, albeit small, numerical uncertainty.

We first construct the finite, renormalized partonic coefficient function through N$^4$LO in QCD using the definitions of the previous sections. We define
\beq
\eta_{ij,\,\text{thr.}}^{B}( z,\mu^2)=\lim\limits_{ z\to 1}\eta_{ij}^{B}( z,\mu^2)=H_{ij}^B(
\mu^2)  S^{R_{ij}}_{\text{thr.}}(z,\mu^2).
\eeq
We can then compare this with the existing results of ref.~\cite{Das:2020adl} and find agreement. 
Next, we want to exploit that the soft function is described by Wilson lines, as described above. 
This implies that the difference between the threshold soft function determined from the DY and Higgs boson cross section is only given by the color representation of the Wilson lines associated with the ingoing partons. 
To make this statement manifest we take the logarithm of the soft function and replace color factors as indicated below.
\bea
\label{eq:casimirscaling}
&&\log\left(\eta_{ij,\,\text{thr.}}^{DY}( z,\mu^2)\Big/ H_{ij}^{DY}(\mu^2)  \right)\\
&&\stackrel{!}{=}\frac{C_F}{C_A}\log\left(\eta_{ij,\,\text{thr.}}^{Higgs}( z,\mu^2)\Big/ H_{ij}^{Higgs}(\mu^2)  \right)\Bigg|_{C^4_{AA}\to C_A/ C_F C^4_{AF},C^4_{AF}\to C_A/ C_F C^4_{FF}}+\mathcal{O}(\alpha_S^5),\nonumber
\eea
where $C^{4}_{R_1R_2}$ denote quartic Casimir operators (see appendix~\ref{sec:color}).
The principle governing the above identity is often referred to as generalized Casimir scaling~\cite{Moch:2018wjh}.

We use eq.~\eqref{eq:casimirscaling} to constrain some of the currently unknown coefficients of the third order splitting function or to derive relations among them. 
In refs.~\cite{Das:2020adl,Das:2019btv} the unknown coefficients of the $\gamma_f^{r,(4)}$ (half the coefficient of $\delta(\bar z)$ in the four loop quark and gluon splitting function) are organized in terms of the the contributing color factors. 
For example,
\beq
	\frac{1}{2}\gamma_f^{\text{fundamental},(4)}=C_F^4 b^4_{q,\,C_F^4}+C_F^3C_A b^4_{q,\,C_F^3C_A}+n_f C_F^3 b^4_{q,\,C_F^3n_f}+\dots.
\eeq
In the end we are able to determine all but four unknown coefficients for both the gluon and quark splitting function. 
Explicitly we determine the following previously unknown constants:
\bea
b^4_{q,\,C_F^4}&=&-384\zeta_{2}\zeta_{5}-120\zeta_{2}\zeta_{3}+64\zeta_{4}\zeta_{3}-450\zeta_{2}-342\zeta_{4}-2111\zeta_{6}\\
&&+5880\zeta_{7}-2520\zeta_{5}-1152\zeta_{3}^2+2004\zeta_{3}+\frac{4873}{24}.\nonumber\\
b^4_{q,\,C_F^3 C_A}&=&2064\zeta_{2}\zeta_{5}-\frac{1988}{3}\zeta_{2}\zeta_{3}+128\zeta_{4}\zeta_{3}+1167\zeta_{2}\nonumber\\
&&+2167\zeta_{4}+\frac{79297\zeta_{6}}{18}-10920\zeta_{7}-976\zeta_{5}+3220\zeta_{3}^2-3260\zeta_{3}-\frac{2085}{4}.\nonumber\\
b^4_{q,\,C_F^2 C_A^2}&=&-2104\zeta_{2}\zeta_{5}+\frac{2096}{9}\zeta_{2}\zeta_{3}-32\zeta_{4}\zeta_{3}-\frac{46771\zeta_{2}}{27}\nonumber\\
&&-\frac{60850\zeta_{4}}{27}-\frac{5497\zeta_{6}}{2}+8610\zeta_{7}+\frac{5354\zeta_{5}}{9}-\frac{7102\zeta_{3}^2}{3}+\frac{129662\zeta_{3}}{27}+\frac{29639}{36}.\nonumber
\eea
Furthermore, we find the following relations.
\bea
b^4_{g,\,n_fC_F^2C_A}&=&b^4_{q,\,n_f C_FC_A^2}.\\
b^4_{g,\,C^4_{AA}}&=&b^4_{q,\,C^4_{AF}}-272\zeta_{2}\zeta_{3}+\frac{1184\zeta_{2}}{3}-\frac{508\zeta_{4}}{3}+\frac{748\zeta_{6}}{9}\nonumber\\
&&+\frac{760\zeta_{5}}{3}-\frac{784\zeta_{3}}{3}-\frac{800}{9}.\nonumber\\
b^4_{g,\,C_A^4}&=&-\frac{1}{24} b^4_{q,\,C^4_{AF}}+80\zeta_{2}\zeta_{5}-\frac{3902}{9}\zeta_{2}\zeta_{3}+168\zeta_{4}\zeta_{3}+\frac{2098\zeta_{2}}{27}+\frac{8965\zeta_{4}}{54}\nonumber\\
&&-\frac{19129\zeta_{6}}{54}+700\zeta_{7}-\frac{14617\zeta_{5}}{9}+\frac{682\zeta_{3}^2}{3}+\frac{48088\zeta_{3}}{27}+\frac{50387}{486}.\nonumber
\eea
The identified coefficients and relations are consistent with the numerical values found in refs.~\cite{Das:2020adl,Das:2019btv}.
The four remaining coefficients are known only numerically as determined by table 1 of ref.~\cite{Das:2019btv} and we show their values here.
\bea
b^4_{q,\,n_fC_F^2C_A}&=&-455.247 \pm 0.005.\\
b^4_{q,\,C^4_{AF}}&=&-998.0 \pm 0.2.\nonumber\\
b^4_{q,\,C^4_{FF}}&=&-143.6 \pm 0.2. \nonumber\\
b^4_{q,\,n_fC_F^3}&=&80.780 \pm 0.005. \nonumber
\eea

Finally, we are able to determine the threshold anomalous dimension analytically up to the four unknown coefficients. 
We define the perturbative expansion of the threshold anomalous dimension as
\beq
\gamma_{\text{thr.}}^{r}=
\sum\limits_{i=0}^\infty\left(\frac{\alpha_S(\mu^2)}{\pi}\right)^i\gamma_{\text{thr.}}^{r,(i)}.
\eeq
Numerically, we find that 
\beq
\gamma_{\text{thr.}}^{\text{fundamental},(4)}=-41.8 \pm 0.01\%,\hspace{1cm}
\gamma_{\text{thr.}}^{\text{adjoint},(4)}=-114.964 \pm 0.04\%.
\eeq
We include the threshold anomalous dimension and the threshold soft function in analytic form as electronically readable files together with the arXiv submission of this article.

\subsection{Thrust and $N$-jettiness anomalous dimensions to N$^4$LO}
Consistency relations among SCET factorization theorems allow us to relate the threshold anomalous dimension to the the anomalous dimensions driving the logarithmic behavior of $\rm{SCET}_I$ observables, which is a large class of observables including DIS at large$-x$, the jet-mass observable, as well as the thrust, $C$-parameter, and $N$-jettiness \cite{Stewart:2010tn} event shapes.

Given the universality of $\rm{SCET}_I$ anomalous dimensions \cite{Stewart:2010qs}, it suffices to find a relation for one of these observables. 
For this we take the factorization theorem of the non-singlet structure function at threshold in DIS \cite{Sterman:1986aj,Catani:1989ne,Korchemsky:1993uz,Becher:2006mr}  
\beq \label{eq:DISlargeXfact}
	F_2^{\rm ns}(x,Q^2) =\sum_q e_q^2 Q^2 H(Q^2,\mu^2) \int_x^1 \df \xi J_q\Big(Q^2 \frac{\xi - x}{x}, \mu^2\Big)\fth_q(\xi,\mu^2)\,,
\eeq
where $J_q$ is the $\rm{SCET}_I$ jet function that also appears in the factorization theorem for thrust \cite{Becher:2008cf,Abbate:2010xh,Abbate:2012jh} and $N$-Jettiness \cite{Stewart:2010tn}. Its RGE reads
\begin{align}\label{eq:JRGE}
	\mu^2 \frac{\df }{\df \mu^2}J_i(s,\mu^2) &= \int \df s^\prime \gamma^i_J(s - s^\prime,\mu^2) J_q(s^\prime,\mu^2)\,,\nn \\
	\gamma^i_J(s,\mu^2) &= -\GammaC^i(\alpha_S(\mu^2))\frac{1}{\mu^2}\left[\frac{\mu^2}{s}\right]_+ +\frac{1}{2} \gamma^i_J(\alpha_S(\mu^2))\delta(s)\,.
\end{align} 
The RGE invariance of the factorized cross section immediately implies that the non-cusp part of $\gamma^i_J(s,\mu^2)$ is
\begin{align}
	&\gamma^i_J(\alpha_S(\mu^2))=- \gamma_f^i(\alpha_S(\mu^2)) -\gamma_H^i(\alpha_S(\mu^2))\,,
\end{align}
and by using \eq{gammathr} we can rewrite it in terms of the threshold and collinear anomalous dimension%
\footnote{Note also that, since from the thrust factorization it is trivial to show that 
\beq 
	\gamma^i_H(\alpha_S(\mu^2)) + 2\gamma^i_J(\alpha_S(\mu^2)) +  \gamma^i_S(\alpha_S(\mu^2)) =0\,,
\eeq 
with $\gamma^i_S(\alpha_S(\mu^2))$ being the anomalous dimension of the thrust soft function, \eq{SCETIJetRel} implies that the threshold soft function anomalous dimension is the opposite of the thrust soft function anomalous dimension
\beq 
	\gamma^i_S(\alpha_S(\mu^2)) = - \gamma_{\text{thr.}}^i(\alpha_S(\mu^2)) \,.
\eeq
}%
\begin{align} \label{eq:SCETIJetRel}
	2\gamma^i_J(\alpha_S(\mu^2))=  \gamma_{\text{thr.}}^i(\alpha_S(\mu^2))-\gamma^i_H(\alpha_S(\mu^2))\,.
\end{align}
For the $\rm{SCET}_I$ beam function \cite{Stewart:2010qs} $B_i(t,z,\mu^2)$, one can either use the equivalence between the $\rm{SCET}_I$ jet and beam function anomalous dimensions and \eq{SCETIJetRel}, or repeat this exercise using the generalized threshold factorization theorem \cite{Lustermans:2019cau}, to show that the following relation holds
\begin{align}
	\mu^2 \frac{\df }{\df \mu^2} B_i(t,z,\mu^2) &=\int \df t^\prime \gamma^i_B(t - t^\prime,\mu^2) B_i(t^\prime,\mu^2)\,, \\
		\gamma^i_B(t,\mu^2) &= -\GammaC^i(\alpha_S(\mu^2))\frac{1}{\mu^2}\left[\frac{\mu^2}{t}\right]_+ +\frac{1}{4}\underbrace{\Big( \gamma_{\text{thr.}}^i(\alpha_S(\mu^2))-\gamma^i_H(\alpha_S(\mu^2))\Big)}_{2 \gamma^i_B(\alpha_S(\mu^2))}\delta(t)\,.\nn
\end{align}
For completeness we include the jet/beam function anomalous dimension in analytic form as electronically readable files together with the arXiv submission of this article.

\section{Conclusions}
\label{sec:conclusions}
Thoughout this paper we computed analytic results for so-called soft master integrals. 
These integrals play a crucial part in the analytic computation of scattering cross sections involving two identified hadrons, and we discussed how these integrals relate to LHC production cross sections, semi-inclusive deeply inelastic scattering and $e^+e^-$ annihilation. 
Our integrals are essential ingredients to compute perturbative cross sections through N$^3$LO and  beyond in QCD and QED perturbation theory. 

We have presented explicit analytic result for differential and inclusive soft master integrals as a Laurent series in the dimensional regulator for partonic scattering processes involving two initial state partons and two or three soft final state partons on top of a colorless final state. 
Our calculation extends available results in the literature, as we include the first nine terms in the expansion in the dimensional regulator. 

Our differential master integrals are integrated analytically over the final state parton momenta and retain all differential dependence on the four momentum of the colorless final state. 
These integrals depend in a non-trivial fashion on one dimensionless variable in terms of functions expressed as harmonic polylogarithms.
Our inclusive soft master integrals are in addition integrated over the degrees of freedom of the colorless final state particle and are given by linear combinations of multiple zeta values. 
Our differential and inclusive master integrals are so-called pure functions of uniform transcendental weight.
We discuss explicitly how the computation of such soft master integrals is greatly facilitated by the simultaneous computation of differential and inclusive soft master integrals in conjunction with the use of the method of differential equations.

We build on a prior computation of the inclusive cross section for the production of a Higgs boson or a lepton pair at the LHC at the production threshold and express the corresponding partonic coefficient function in terms of our soft master integrals. 
With this we compute threshold corrections to these partonic cross sections to two higher powers in the dimensional regulator. 
These results form a crucial ingredient for a future determination of the Drell-Yan and Higgs boson production cross section at N$^4$LO in perturbative QCD.

Finally, we recap the factorization of cross sections describing the production of a colorless final state in the threshold limit. 
We extract all required anomalous dimensions and find agreement with previous results. 
We then explicitly determine the so-called threshold soft function at N$^4$LO in perturbative QCD up to one constant that is yet to be determined from a genuine computation at this order.
Furthermore, we extract the threshold anomalous dimension through fourth loop order. 
Finally, using the generalized Casimir scaling property of the threshold soft function, we obtain new analytic results for several coefficients in the four loop Altarelli-Parisi splitting functions in the term proportional to $\delta(1-z)$.

We present many of our results as ancillary files in electronically readable form appended to the arXiv submission of this article, and we enumerate them here:
\begin{enumerate}
\item Definitions and solutions for canonical inclusive soft master integrals for RR, RRV and RRR production cross sections. In our solutions we set $\bar z=Q^2=1$ as the functional dependence on this variables is easily restored by multiplying the solutions with $(\bar z Q)^{-2(m+L)\eps}$, where $m$ and $L$ are the number of soft partons and loops respectively.
\item Definitions and solutions for canonical differential soft master integrals for RR, RRV and RRR production cross sections. In our solutions we set $w_1=w_2=Q^2=1$ as the functional dependence on this variables is easily restored by multiplying the solutions with $(w_1 w_2 Q^2)^{-(m+L)\eps}$, where $m$ and $L$ are the number of soft partons and loops respectively. 
Furthermore, we include the matrices A for the canonical differential equations of our differential soft master integrals $I$, which take the form
\beq
\df \vec{I}=\df A \cdot \vec{I}.
\eeq
\item The bare inclusive soft-virtual cross section for the production of a Higgs boson or a Drell-Yan lepton pair through N$^3$LO in perturbative QCD and including two additional powers in the dimensional regulator, i.e., in total the first nine terms in the expansion in $\epsilon$ at every perturbative order.
\item The renormalized, finite soft function (see eq.~\eqref{eq:SFfiniteDef}) for the production of a colorless final state by scattering of quarks of gluons through N$^4$LO in perturbative QCD. 
The soft function is determined up to one remaining constant at N$^4$LO multiplying a Dirac delta distribution of $\bar z$.
\item The threshold anomalous dimension through N$^4$LO in perturbative QCD. In particular, we include $\gamma_{\text{thr.}}^{r}(\alpha_S(\mu^2))$ of eq.~\eqref{eq:gammathr}.
\item The N-Jettiness beam function anomalous dimension $\gamma^r_J(\alpha_S(\mu^2))$ through N$^4$LO in perturbative QCD of eq.~\eqref{eq:SCETIJetRel}.
\end{enumerate}

\section*{Acknowledgments}
GV and BM are supported by the United States Department of Energy, Contract DE-AC02-76SF00515. 

\appendix
\section{Ultraviolate Renormalisation}
\label{sec:uvren}
The strong coupling constant renormalises as
\beq
\alpha_S=\alpha_S(\mu^2)(4\pi)^{-\epsilon}e^{\epsilon \gamma_E} Z_\alpha(\mu^2),
\eeq
with
\bea
Z_{\alpha_S}(\mu^2)&=&1
+a_S(\mu^2)\left[-\frac{\beta_0}{4\epsilon}\right]\\
&+&a^2_S(\mu^2)\left[\frac{\beta_0^2}{16\epsilon^2}-\frac{\beta_1}{16\epsilon}\right]\nonumber\\
&+&a^3_S(\mu^2)\left[\frac{-\beta_0^3}{64\epsilon^3}+\frac{7\beta_1\beta_0}{384\epsilon^2}-\frac{\beta_2}{192\epsilon}\right]\nonumber\\
&+&a^4_S(\mu^2)\left[\frac{\beta_0^4}{256\epsilon^4}-\frac{23\beta_1\beta_0^2}{3072\epsilon^3}+\frac{3\beta_1^2}{2048\epsilon^2}-\frac{5\beta_0\beta_2}{1536\epsilon^2}-\frac{\beta_3}{1024\epsilon}\right]+\mathcal{O}(a_S^4(\mu^2)),\nonumber
\eea
with $a_S(\mu^2)=\alpha_S(\mu^2)/\pi$ and $\beta_i$ the coefficients of the QCD beta function~\cite{Tarasov:1980au,Larin:1993tp,vanRitbergen:1997va,Czakon:2004bu,Baikov:2016tgj,Herzog:2017ohr}.
The Wilson coefficient~\cite{Chetyrkin:1997un,Kramer:1996iq,Schroder:2005hy,Chetyrkin:2005ia} for the heavy top effective theory renormalizes as 
\beq
C_t=Z_t(\mu^2) C_t(\alpha_S(\mu^2),m_t^2,\mu^2),\hspace{1cm}Z_t(\mu^2)=\frac{1}{1-\frac{\beta(\alpha_S(\mu^2))}{\epsilon}}.
\eeq
\section{Mass Factorisation}
\label{sec:convolute}
The mass factorization counter term absorbs collinear singularities into a suitable redefinition of the parton distribution functions. 
It is defined in terms of the the following differential equation
\beq
\label{eq:DGLAP}
\partial_{\mu^2}\Gamma_{ij}(x,\mu^2)=-a_S(\mu^2) \Gamma_{ik}(x,\mu^2) \circ_x P_{kj}(x,\mu^2), \hspace{1cm}\partial_{\mu^2}=\frac{\df}{\df \log (\mu^2)}.
\eeq
Here, $P_{ij}$ are the Altarelli-Parisi splitting functions~\cite{Moch:2004pa,Vogt:2004mw,Ablinger:2014nga,Ablinger:2017tan,Blumlein:2021enk} and the above differential equation is derived from DGLAP evolution of PDFs~\cite{Gribov:1972ri,Altarelli:1977zs,Dokshitzer:1977sg}.
We note that 
\beq
\partial_{\mu^2} a_S(\mu^2)=-\epsilon a_S(\mu^2)-\beta(a_S(\mu^2)).
\eeq
We expand $\Gamma_{ij}(x,\mu^2)$ in the strong coupling constant and define 
\beq
\Gamma_{ij}(x,\mu^2)=\sum_{o=0}^\infty a_s(\mu^2)^o\, \Gamma^{(o)}_{ij}(x,\mu^2),\hspace{1cm}\Gamma_{ij}^{(0)}(x,\mu^2)=\delta_{ij}\delta(1-x).
\eeq
In eq.~\eqref{eq:DGLAP} we make use of the Mellin convolution
\beq
f(z)\circ_z g(z)=\int_z^1 \frac{\df x}{x} f(x) g\left(\frac{z}{x}\right).
\eeq
In the main part of this article we are interested in the soft limit of functions which enter such Mellin convolutions. 
A convolution of two functions $f(x)$ and $g(x)$ which have both been computed in the limit $x\to1$ will introduce power suppressed terms in $(1-x)$.
It is thus useful to introduce another convolution which maintains the correct leading term of the convolution but does not introduce additional power suppressed terms.
\beq
\label{eq:softconv}
f(z)\otimes_z g(z)=\int_z^1 \df x f(x) g\left(1-x+z\right).
\eeq
The above definition is easily found by expanding the original Mellin transform around the limit of $z\to 1$.
We also define the mass factorization counter term in this limit to be 
\beq
\Gamma^{\text{adjoint}}(x,\mu^2)=\lim_{x\to 1}\Gamma_{gg}(x,\mu^2),\hspace{1cm}
\Gamma^{\text{fundamental}}(x,\mu^2)=\lim_{x\to 1}\Gamma_{qq}(x,\mu^2).
\eeq
\section{Color}
\label{sec:color}
We define
\beq
C_A=n_c,\hspace{1cm}C_F=\frac{n_c^2-1}{2n_c},
\eeq
where $n_c$ is the number of colors.
We define the Casimir values by
\beq
C^n_{R_1 R_2}=\frac{1}{(n!)^2 }\left(T^{\{a_1}_{R_1}\dots T^{a_n\}}_{R_1}\right)\left(T^{\{a_1}_{R_2}\dots T^{a_n\}}_{R_2}\right),
\eeq
where the curly brackets indicate the fully symmetric trace over the terms and $T_R^a$ is a generator of $SU(n_c)$ in representation $R$.
We find for
\begin{itemize}
\item $n=3:$
\beq
C^3_{FF}=\frac{1}{16}\frac{n_c^2-4}{n_c}(n_c^2-1),\hspace{1cm}
C^3_{FA}=0,\hspace{1cm}
C^3_{AA}=0.
\eeq
\item $n=4:$
\bea
C^4_{FF}&=&\frac{1}{96}\frac{18-6 n_c^2+n_c^4}{n_c^2}(n_c^2-1),\\
C^4_{FA}&=&\frac{1}{48}n_c(6+n_c^2)(n_c^2-1),\nonumber\\
C^4_{AA}&=&\frac{1}{24}n_c^2 (36+n_c^2)(n_c^2-1).\nonumber
\eea
\end{itemize}

\addcontentsline{toc}{section}{References}
\bibliographystyle{jhep}
\bibliography{refs}

\end{document}